\documentstyle[12pt,twoside]{article}
 \font\twelve=cmbx10 at 15pt
\font\ten=cmbx10 at 12pt

\renewcommand{\thefootnote}{\fnsymbol{footnote}}

\begin{document}

\begin{titlepage}

\begin{center}

{\ten Centre de Physique Th\'eorique\footnote{Unit\'e Propre de
Recherche 7061} - CNRS - Luminy, Case 907}

{\ten F-13288 Marseille Cedex 9 - France }

\vspace{2 cm}

{\twelve AUTOMATIC BIASES CORRECTION}

\vspace{0.3 cm}
\setcounter{footnote}{0}
\renewcommand{\thefootnote}{\arabic{footnote}}

{\bf Roland TRIAY\footnote{and Universit\'e de Provence, Marseille}}

\vspace{2 cm}

{\bf Abstract}

\end{center}

The key point limits to define the
{\it statistical model} describing
the data distribution. Hence, it turns out that the characteristics
related to the so-called
Inverse Tully-Fisher relation and the Direct relation are maximum
likelyhood ({\sc ml})
estimators of different statistical models, and we obtain coherent
distance estimates as long as
the same model is used for the calibration of the TF relation and
for the determination of
distances. The choice of the model is motivated by reasons of {\em
robustness} of statistics,
which depends on selection effects in observation.
\vspace{2 cm}


\bigskip

\noindent October 1994

\noindent CPT-94/P.3082

\bigskip

\noindent anonymous ftp or gopher: cpt.univ-mrs.fr

\end{titlepage}

\noindent 1. {\large INTRODUCTION}
\vspace{0.5cm}

\noindent The method of correcting biases in estimating the
distances of galaxies is one of the
major problem which must be solved for a better understanding of
the cosmic velocity fields, see
\cite{GouEtal89,LynEtal88,Tee90} and P.Teerikorpi~(this
conference). If one keeps in mind that
any technique of fitting is intimately related to a {\em
statistical model} \cite{BigTri90} then
one understands that the cause of the weak convergence of present
debates, for arguing on the use
of either the direct Tully-Fisher relation ({\rm DTF}) or the
inverse relation ({\rm ITF}),
interprets as an {\em unsufficiently handled formulation} of the
problem. The obstacle toward a
consensus can be overcomed by arguing on the model instead of the
technique of fitting. Most of
the present contribution is a brief presentation of results
obtained in \cite{TriayEtal94a}.

\vspace{1cm}
\noindent 2. {\large BASICS OF THE BIASES CORRECTION}
\vspace{0.5cm}

\noindent To ask oneself whether the statistical estimator ({\em
statistic}) corresponds to the
model parameter for which it has been made up, is indeed a sensible
question. Generically, a
statistic $\hat{\theta}$ of a given parameter $\theta$ provides us
with an estimate
\begin{equation}\label{Statistic}
\hat{\theta}_{N} = \theta + \epsilon_{N}
\end{equation}
within a (unkown) random error $\epsilon_{N}$, where $N$ denotes
the sample size. Thus, the
accuracy of such an estimate can be discussed only in terms of
characteristics describing the
probability law of $\epsilon_{N}$. For example, it is clear that
the smaller the variance of
$\epsilon_{N}$ the more precise such an estimate, as long as it is
not biased. By definition,
``$\hat{\theta}_{N}$ is biased when the {\em expected value} of
$\epsilon_{N}$ is
not zero''. While an unbiased statistic shows a smaller variance,
it turns out that
such a property is not essential, it can be reached asymptotically
(i.e., for
$N\rightarrow\infty$).

Actually, the typical problem of biases in the present fields of
interest is intimately related
to the question of whether the selection effects in observation are
correctly taken into account
in the statistical model. In other words, we easily understand that
one can obtain unbiased
statistics as long as the {\em probability density} ({\em pd})
describing the
$\epsilon_{N}$-distribution is known, which requires a
``statistical modeling'' of the data. At
this point, which is the first step toward the understanding of any
problem involving
observations, nothing prevents us to use solely the {\it maximum
likelihood} ({\sc ml}) technique
for obtaining suitable statistics. The enormous advantage of such
an approach is to provide us
unambiguously with a unique fitting technique, which prevents us
{}from subjective speculations on
diagrams.

\vspace{1cm}
\noindent {\it 2.1 The Statistical Model -- The Method}
\vspace{0.5cm}

\noindent The {\em pd} describing the distribution of observables
reads
\begin{equation}\label{Model}
dP_{\rm obs} = \frac{\phi}{P_{\rm th}(\phi)}~dP_{\rm th},
\end{equation}
where $0\leq\phi\leq1$ is a {\em selection function} in
observation, $dP_{\rm th}$ describes the
distribution of intrinsic variables related to sources and $P_{\rm
th}(\phi) = \int{\phi~dP_{\rm
th}}$ is the normalization factor. Obviously, {\em working
hypotheses} are required in order to
define the selection function $\phi$ (in term of observables) and
the theoretical {\em pd}
$dP_{\rm th}$ (in term of intrinsic quantities). Hence, we can
write the likelihood
function\footnote{Actually, it is more convenient to use its
natural logarithm.} ${\cal L}_{\rm
obs} = {\cal L}_{\rm th}-\ln\left(P_{\rm th}(\phi)\right)$, where
${\cal L}_{\rm th}$ corresponds
to the {\em pd} $dP_{\rm th}$, and the {\sc ml} statistic is
derived from the equation
\begin{equation}\label{MLEquations}
\partial_{\theta}{\cal L}_{\rm obs} = 0.
\end{equation}
Note the feature which informs on the presence of biases~: a
$\theta$-statistic related to
equation $\partial_{\theta}{\cal L}_{\rm th} = 0$ differs from
$\hat{\theta}_{N}$ if
$\partial_{\theta}P_{\rm th}(\phi)\neq 0$.

If the sample is not peculiar then the {\sc ml} statistic
$\hat{\theta}_{N}$
provides us with the {\em most probable value} of $\theta$ within a
given {\em accuracy},
althought it is not necessarely unbiased. For recovering an
accurate estimate, the {\sc ml}
statistic must be shifted by the expected value of $\epsilon_{N}$,
\begin{equation}\label{StatisticApriori}
\theta \approx \hat{\theta}_{N}-P_{\rm
obs}\left(\epsilon_{N}\right),
\end{equation}
while (in practice) such an approach might demand cumbersome
calculations. However, according to
the {\em Central limit theorem}, if $N$ is large enough then one
expects that the discrepancy is
neglectable ($\epsilon_{N}\approx 0$), which means that the {\sc
ml} statistic is asymptotically
unbiased. Finally, we easily understand that any result is
warranted as long as the distribution
of variables involved in the calculation is correctly described by
such a model.

The calculation of the {\em mean absolute magnitude} of galaxies
{}from a magnitude limited sample
is a pedagogic example for comparing the {\sc ml} approach to the
Malmquist~(1920) calculation
\cite{Mal20}. The statistical model is based on --~a Gaussian
luminosity distribution function;
--~a uniform spatial distribution; --~and a sharp cutoff at a
limiting magnitude $m_{\rm lim}$.
Thus $ dP_{\rm th} \propto g_{\rm
G}(M;M_{\circ},\sigma_{M})dM~e^{\beta\mu}d\mu$, where
$\beta=3\ln10\,/5$, and the selection function $\phi_{m}(m) =
\theta\left(m_{\rm lim} - (\mu +
M)\right)$,  where $\theta$ denotes the Heaveside distribution
function. Since the normalization
factor $P_{\rm th}\left(\phi_{m}\right) \propto
\exp\left(\frac{\beta}{2}\sigma_{M}^{2}-M_{\circ}\right)$ depends
on $M_{\circ}$, the standard
statistics are expected to be biased. Indeed, if $\sigma_{M}$ is
unknown then the {\sc ml}
equations provide us with the following system of unbiased
statistics
\begin{eqnarray}\label{LikeStat} M_{\circ} &=& \langle M \rangle +
\beta\sigma_{M}^{2},\\
\sigma_{M}^{2} &=&
\frac{1}{2\beta^{2}}\left(\sqrt{1+4\beta^{2}\langle(M-M_{\circ})^{2}\rangle}
-
1\right), \end{eqnarray}
which can be solved by Newton's method. Note that the {\sc ml}
approach generalizes the
Malmquist~(1920) solution.

\vspace{1cm}
\noindent 3. {\large ABOUT THE DISTANCE ESTIMATE OF GALAXIES}
\vspace{0.5cm}

\noindent The goal is to estimate a distance modulus from the
observed apparent magnitude
$m=M+\mu$ and the distance estimator $p$, which gives a rough
estimate of the absolute magnitude
$M \approx a.p+b$ by means of the Tully-Fisher relation (for
spirals) \cite{TulFis77}, or the
Faber-Jackson relation (for ellipticals) \cite{FabJac76}. The
distribution of intrinsic
quantities is described by $dP_{\rm th} =
\kappa(\mu)d\mu~F(p,M)dpdM$, where $\kappa(\mu)$
accounts for the galaxies distribution in space and $F(p,M)$ for
the distribution in the $p$-$M$
plane\footnote{It must be noted that this distribution is different
{}from the one in the
TF-diagram, which is described by a {\it pd} $\propto F(p,M)dpdM
\int_{\mu} \phi
\kappa(\mu)d\mu$.}. For reasons that become clear in the following,
we describe the $p$-$M$
distribution according to different statistical models
\begin{eqnarray}\label{ModelTF}
F(p,M)dpdM = g_{\rm G}(\zeta;0,\sigma_{\zeta})d\zeta \times \left\{
\begin{tabular}{ll}
$f_{M}(M;M_{\circ},\sigma_{M})dM$  & (ITF)
\\ $f_{p}(p;p_{\circ},\sigma_{p})dp$ & (DTF)
\end{tabular}
 \right.,
\end{eqnarray}
where $\zeta=a.p+b-M$ accounts for the {\em intrinsic} dispersion
about the TF-relation, it is
assumed to be Gaussian distributed about zero and with standard
deviation $\sigma_{\zeta}$.

Table~\ref{CalibrationStat1} gives the related {\sc ml} statistics
of parameters $a$, $b$ and
$\sigma_{\zeta}$ in term of statistics of the covariance (Cov), the
standard deviation
($\Sigma$), the mean ($\langle.\rangle$) and the correlation
coefficient ($\rho$).
\begin{table}[h]\caption []{Calibration Statistics
}\label{CalibrationStat1} \begin{center}\begin{tabular}{|c|cc|}
     \hline
          & ITF        & DTF   \\
      \hline
       $a$     & $\Sigma(M)^2/{\rm Cov}(p,M)$& ${\rm
Cov}(p,M)/\Sigma(p)^{2}$\\
        $b$     & $\langle M \rangle - a \langle p \rangle$&
$\langle M \rangle - a \langle p \rangle + \beta
\sigma_{\zeta}^{2}$ \\
        $\sigma_{\zeta}$ & $|\rho(p,M)|^{-1} \Sigma(M)\sqrt{1-
\rho^2(p,M)}$ & $\Sigma(M)
\sqrt{1- \rho^2(p,M)}$\\
        $Constraints$ & $\phi_{p}=1$ & $\left\{
\begin{array}{c}
\phi_{p}\phi_{\mu}=1\\
\phi_{m}(m) = \theta(m_{\rm lim}-m)\\
\kappa(\mu) \propto \exp(\beta \mu)\\
f_{p}(p) = g_{\rm G}(p;p_{\circ},\sigma_{p})
\end{array}
\right.$\\
     \hline
\end{tabular}\end{center}
\end{table}
It is then clear that the identifications of $a$ to the ``slope''
and $b$ to the ``zero-point''
of the TF relation are model dependent. These statistics are valid
as long as the working
hypotheses ({\em Constraints}) are fulfilled, in particular the
absence of $p$-selection effects.
They must be corrected for a bias due to measurement errors, which
also increase the dispersion.
However, for typical samples, we obtain
estimates with a relative (1~$\sigma$) accuracy of 7\% for $a$ and
15\% for $b$. The simulations
show that the main source of error is actually due to the small
size of the calibration sample
($\approx 30$ galaxies) instead of errors.

Hence, we understand that the choice of the model must be discussed
as a {\em strategy}. Indeed,
the {\rm ITF} model is much less constraintfull than the {\rm DTF},
which makes the related
statistics more {\em robust} (see e.g. \cite{Hendry90}). In the
other hand, one might expect that
(in general) the more numerous the working hypotheses the more
precise the related statistic, the
simulations show that the accuracy increases of 5\% in the DTF
model. However, it is clear that
if one of these hypotheses is not so correct then the estimate is
bogus. In practice, such a
characteristic forces us to prefer the {\rm ITF} approach, because
of the usual conditions in
observation. Nevertheless, it turns out that both models show the
same robustness if they are
improved for taking into account $p$-selection effects (in prep.).

In order to estimate a likely distance modulus $\mu$ of a galaxy
{}from the same statistical model
we have to assume that the galaxy belongs to {\em the same
population} of the calibration sample.
According to the Bayesian schema\footnote{It is prefered to the
{\em frequentist} schema
\cite{Hendry90} because the sample has a unique element, $\mu$
interprets as a model parameter of
the {\it pd} $dP_{\rm obs}( m_{k},p_{k}\mid\mu)$.}, provided the
observables $m=m_{k}$ and
$p=p_{k}$, the distribution of possible outcomes reads $dP_{\rm
obs}(\mu\mid
m_{k},p_{k})\propto\int_{M}\int_{p}\delta(m-m_{k})\delta(p-p_{k})dP_{\rm
obs}$, which gives
\begin{eqnarray}\label{distance}
f_{\mu}(\mu;\mu_{\circ}^{(k)},\sigma_{\mu}^{(k)}) d\mu
&\propto& \kappa(\mu)\,g_{\rm
G}(\mu;\tilde{\mu}_{k},\sigma_{\zeta}) d\mu\times\left\{
\begin{tabular}{ll} $f_{M}(m_{k} - \mu;M_{0},\sigma_{M})$ & (ITF)
\\ $1$ & (DTF) \end{tabular}
 \right.\nonumber
\end{eqnarray}
where $\tilde{\mu}_{k} = m_{k} - (a.p_{k} + b)$ is model dependent,
the mean $\mu_{\circ}^{(k)}$
and the standard deviation $\sigma_{\mu}^{(k)}$ depend on working
hypotheses which specify the
functions $\kappa$ and $f_{M}$. The value $\mu_{\circ}^{(k)}$
interprets as an unbiased estimate
of the distance modulus. The difference between $\mu_{\circ}^{(k)}$
and $\tilde{\mu}_{k}$ is not a
bias of Malmquist type but a {\em volume correction}, since the
Dirac's distribution functions
cancel the dependence of any selection function on $m$ and on $p$.
Finally, it is important to
mention that if the distribution function $f_{\mu}$ is not symetric
about $\mu_{\circ}^{(k)}$
then this unbiased distance estimate does not necessarely
correspond to the {\em most probable
distance} \begin{equation}\label{MLDistance}
\breve{\mu}_{k}=\tilde{\mu}_{k}+\sigma_{\zeta}^{2}\partial_{\mu}\ln
\kappa(\mu)+\sigma_{\zeta}^{2}
\left\{ \begin{tabular}{ll} $\partial_{\mu}\ln f_{M}(m_{k} -
\mu;M_{0},\sigma_{M})$ & (ITF)
\\ $0$ & (DTF)
\end{tabular}
 \right.,
\end{equation}
which is defined as the root of equation
$\partial_{\mu}f_{\mu}(\mu;\mu_{\circ}^{(k)},\sigma_{\mu}^{(k)}) =
0$. Therefore, we see that the
problem of the distance estimate of individual galaxies depends on
the choise of the ``strategy of
gambling'' (i.e., either one minimizes the random error or one bets
to the most likely value
within a given accuracy). According to Eq.\,(\ref{distance}), it is
important to note that the
{\rm DTF} statistic does not require information on the luminosity
distribution function, which
makes the related distance estimate more robust than the {\rm ITF}
one. Therefore, we understand
that if $p$-selection effects are absent then it is more convenient
to use the ITF model for the
calibration step, while the DTF model is prefered for the distance
estimate. The possibility to
get benefit of both advantages is presented by S. Rauzy~(this
conference).

If $f_{\rm M}=g_{\rm G}$ and $\kappa(\mu)\propto e^{\beta\mu}$ then
the distance estimates
coincide,
\begin{equation}\label{MLDistanceEx} \breve{\mu}_{k} =
\mu_{\circ}^{(k)} = \left\{\begin{tabular}{ll}
$\frac{1}{1+\gamma^{2}}
\left(\left(\tilde{\mu}_{k}+\beta\sigma_{\zeta}^{2}\right)  +
\gamma^{2}(m_{k} - M_{\circ})
\right)$ & (ITF) \\
 $\tilde{\mu}_{k}+\beta\sigma_{\zeta}^{2}$ & (DTF)
\end{tabular}
 \right.
\end{equation}
where $\gamma = \sigma_{\zeta}^{\rm ITF}/\sigma_{M}$ is a tiny
quantity. The formal comparison of
statistics shows that the discrepancy is a random variable of zero
mean and neglectable standard
deviation. Moreover, if the estimation of the mean $M_{\circ}$
limits to the calibration sample
then both models provide us with {\em the same distance
estimate}\footnote{Since we have the {\sc
ml} estimate $M_{0} = a^{\rm ITF}\langle p \rangle_{1} + b^{\rm
ITF} + \beta(\Sigma_{1}(M))^2$.}.

\end{document}